\documentstyle[12pt]{article}

\begin{document}
\baselineskip=1.2\baselineskip
\renewcommand{\thefootnote}{\fnsymbol{footnote}}

\setcounter{equation}{0}
\setcounter{section}{0}
\renewcommand{\thesection}{\arabic{section}}
\renewcommand{\theequation}{\arabic{equation}}

\pagestyle{plain}
\begin{titlepage}

\begin{flushright}
SPhT/94-065 \\ hep-ph/9406318
\end{flushright}
\vskip 2.0cm
\begin{center}
{\large{\bf  {The Free Energy of Hot QED at Fifth Order\\}}}
\end{center}
\vskip 1.2cm
\begin{center}
{Rajesh R. Parwani\footnote{email : parwani@wasa.saclay.cea.fr}}
\end{center}
\vskip 0.5cm
\centerline{Service de Physique Th{\'e}orique, CE-Saclay}
\centerline{ 91191 Gif-sur-Yvette, France.}
\vskip 2.0cm
\centerline{PACS 12.20.Ds, 52.60.+h, 11.15.Bt, 12.38.Mh.}

\centerline{12 May 1994 }

\vskip 3.0 cm
\centerline{\bf Abstract}
\vskip 0.2cm
The order $e^5$ contribution to the pressure of massless
quantum electrodynamics at
nonzero temperature is determined explicitly.
An identity is also
obtained relating a gauge-invariant piece
of the pressure at order $e^{2n+3} \ (n \ge 1) $
(from diagrams with only one fermion loop) to the pressure
at order $e^{2n}$.
Prospects for higher order calculations are discussed and
potential
applications  are mentioned. \\

\vfill

\end{titlepage}

\newpage

The three-loop contribution to the equation-of-state of massless
quantum electrodynamics (QED) at temperature $T$ was
obtained recently \cite{CP}, thus extending the well
known two-loop
result \cite{AP}. While three-loop calculations in cold plasmas
($T =0$, but
nonzero chemical potential $\mu$) were done some time
ago \cite{FMB}, same order calculations for hot plasmas have
been delayed by technical
complications due to the presence of statistical distribution
factors
in loop integrals. The three-loop calculation in hot
QED \cite{CP} was
stimulated by in part by a ground breaking three-loop
calculation in hot $\phi^4$ theory \cite{FST}.

The purpose of this letter is to further the results of \cite{CP}
by determining explicitly the next order ($e^5$)
contribution to the pressure of QED with $N$ massless electrons
at nonzero $T$ (but $\mu =0$). Higher orders will also be
discussed
qualitatively. The conventions here are similar to \cite{CP}
and are reiterated for clarity : The imaginary time formalism
is used, whereby the energies are  discrete and imaginary,
$q_0= i n \pi T $, $n$ being an  odd (even) integer for fermions
(bosons). Ultraviolet singularities are regularised
by dimensional
continuation ($4 \rightarrow D$) and renormalisation is
through minimal subtraction.
The wave-vector, $Q_\mu = (q_0,\vec{q})$ lives in a space with
Minkowski metric, $g_{\mu \nu}=$diag($1,-1,-1....-1$) and the
measure for loop integrals is denoted by

\begin{eqnarray}
\int\{dq\}\equiv T\sum_{q_0,odd} \int (dq) \, , \nonumber
\end{eqnarray}

\begin{eqnarray}
\int [dq]\equiv T \sum_{q_0,even}\int (dq) \, , \nonumber
\end{eqnarray}
where
\begin{eqnarray}
\int (dq) = \int {d^{D-1}q\over (2 \pi)^{D-1}} \, .
\end{eqnarray}
Spinor traces are normalised according to  $Tr(\gamma_{
\mu}\gamma_{\nu})=4 g_{\mu\nu}$, and the photon is in the
 Feynman gauge, $ D_{\mu \nu}(Q) = g_{\mu\nu}/Q^2$.

Recall that in a QED plasma static electric fields are
screened with
an inverse screening length given to lowest
order by $e T / \sqrt{3}$.
Consequently a naive perturbative expansion using bare
(unscreened)
propagators produces severe infrared (IR) divergences in
diagrams such
as in Figs.1,2. When these diagrams are resummed one obtains
a finite result
of order $e^3$ \cite{GB,AP} which is nonanalytic in the
coupling $e^2$.
Just as the $e^3$ term is the ``plasmon'' correction
to the two-loop
$e^2$ contribution obtained by dressing the photons, the order
$e^5$ term to be presently calculated
is the plasmon correction to the three-loop $e^4$ result.
The object
that is required in the analysis below is the renormalised,
static, {\it one}-loop
photon polarisation tensor $\Pi_{\mu\nu}(q_0=0,q)$. Gauge
invariance, $Q_{\mu} \Pi^{\mu \nu}(Q) = 0$,
implies $\Pi_{i0}(0,q)=0$
(this is of course true to all orders) while
explicit calculations yield
$\Pi_{ij}(0,q \rightarrow 0) = O(e^2 q^2)$ and
\begin{eqnarray}
\Pi_{00}(0,q) \equiv m^2 + e^2 q^2 h(q) \, ,
\label{burro}
\end{eqnarray}
where $e$ is the renormalised coupling.
The following limits will be required : \\
$ m^2(D \rightarrow 4) = e^2 T^2 N/3$
and
\begin{eqnarray}
h(q=0) = { \mu^{4-D} \ 2 \ N \ (D-2) f_2 \over 3 } \  +
\ {N \over 6 \pi^2 (D-4)} \, .
\label{hippo}
\end{eqnarray}
The second term in (\ref{hippo}) is the UV counterterm and
$\mu$ is the mass scale of dimensional regularisation.
The integral $f_2$ in (\ref{hippo}) is defined through
\begin{eqnarray}
f_n &\equiv&  \int { \{d\,Q\} \over (Q^2)^n} = (2^{2n+1-D}-1)
\, b_n \,
, \nonumber \\
b_n &\equiv& \int  {[d\,Q]\over (Q^2)^n} \,  \nonumber \\
&=&  {2 \ (-1)^n \ T^{D-2n} \ \pi^{{D-1 \over 2}} \over
(2 \pi)^{ 2n} \ \Gamma(n) }
\, \zeta(2n +1 -D) \ \Gamma\left({2n+1-D \over 2}\right)
\, . \nonumber
\end{eqnarray}

Consider now the three-loop,
order $e^4 N$, diagram $G_1$ shown in Fig.3a. It is given by
\begin{eqnarray}
{G_1 \over \mu^{8-2D}} &=& {-e^4 N \over 2} \int { \{ dK \}
[ dQ \ dP ] \
Tr ( \gamma_{\mu} \not\!\! K \gamma_{\alpha}
D^{\alpha \sigma}(P)
(\not\!\! K-\not\!\!P) \gamma_{\sigma}
\not\!\!K \gamma_{\nu} D^{\nu \mu}(Q) (\not\!\!K-\not\!\!Q) )
\over K^4 (K-Q)^2 (K-P)^2 } \, . \nonumber \\
&& \label{giraffe}
\end{eqnarray}

{}From the behaviour of the polarisation tensor discussed above,
it is deduced that when one or more self-energy subdiagrams
are inserted along the photon lines of $G_1$, IR
divergences occur
only for the $\Pi_{00}(0,0) \equiv m^2$ insertions .
These may be resummed into an effective propagator
\begin{eqnarray}
D_{\mu \nu}(Q) &\rightarrow& \left( {g_{\mu \nu} \over Q^2} -
{g_{\mu 0} \
 g_{\nu 0} \ \delta_{q_{0}, 0} \over Q^2} \right) +
{g_{\mu 0} \ g_{\nu 0} \ \delta_{q_{0}, 0} \over Q^2 - m^2}
\; \nonumber \\
&=&  {g_{\mu \nu} \over Q^2} + { m^2 \ g_{\mu 0} \ g_{\nu 0}
\ \delta_{q_{0}, 0}
\over q^2 (q^2 +m^2)}  \nonumber \\
&& \nonumber \\
&\equiv& D_{\mu \nu}(Q) + D^{*}_{\mu \nu}(Q) \, .
\label{rhino}
\end{eqnarray}

When the effective  propagator (\ref{rhino}) is used for the
$Q$- photon in $G_1$  the correction $D^{*}(Q)$
makes the $q$-integral of order $e$,
$ \int {d^3q \ m^2 \over q^2 (q^2 +m^2)} \sim m \sim e$, and
therefore
its  contribution to  $G_1$ is $O(e^5)$. Similarly if both
photons are dressed,
the extra correction to $G_1$ will be of order $e^6$. Thus
the order $e^5$
contribution from dressing $G_1$ is obtained by
keeping one photon
 bare, using the correction $D^{*}$ for the other, and
multiplying the
result by two :
\begin{eqnarray}
{\delta G_1 \over \mu^{8-2D}}= {-e^4 N T } \int \ (dq) { m^2 \
\delta_{q_{0},0} \over q^2 (q^2 + m^2)} \ \int
{ \{dK\} [dP] \ Tr ( \gamma_{0} \not\!\!K \gamma_{\alpha}
(\not\!\!K-\not\!\!P)
\gamma^{\alpha} \not\!\!K
 \gamma_{0}(\not\!\!K-\not\!\!Q) ) \over K^4 P^2 (K-Q)^2
(K-P)^2 } \, .
\nonumber \\
&&
\label{raffe}
\end{eqnarray}
Now scale $\vec{q}=m \vec{x}$ in the above equation. This
results in an
overall external factor $e^4 m \sim e^5$ and since the
$K,P$ integrals are
infrared safe in the $m \rightarrow 0$ limit, one obtains
the exact $e^5$
contribution from $\delta G_1$ as
\begin{eqnarray}
{G_{15} \over \mu^{8-2D}}= {-e^4 N T m^{D-3}} \int
{ (dx) \over x^2 (x^2 + 1)} \
\int { \{dK\} [dP] \ Tr ( \gamma_{0} \not\!\!K \gamma_{\alpha}
(\not\!\!K-\not\!\!P) \gamma^{\alpha} \not\!\!K \gamma_{0}
\not\!\!K ) \over K^6 P^2 (K-P)^2 } \, . \nonumber \\
&& \label{uno}
\end{eqnarray}
Notice that for the $e^5$ calculation, the original complicated
three-loop integral has factorised
completely into the product of a finite integral
and a relatively simple two-loop integral. A similar
analysis for
diagram $G_2$ (Fig.3b) gives for the $e^5$ contribution,
\begin{eqnarray}
{G_{25} \over \mu^{8-2D}} &=& {-e^4 N T m^{D-3} \over 2}
\int { (dx) \over x^2 (x^2 + 1)}
\int { \{dK\} [dP] \ Tr ( \gamma_{0} \not\!\!K \gamma_{\alpha}
(\not\!\!K-\not\!\!P)
\gamma_{0}(\not\!\!K-\not\!\!P) \gamma^{\alpha} \not\!\!K )
 \over K^4 P^2 (K-P)^4 }
\, . \nonumber \\
&& \label{deux}
\end{eqnarray}
For the sum of (\ref{uno}) and (\ref{deux}) I obtain
\begin{eqnarray}
G_{15} + G_{25} &=& -e^4 \ m^{D-3} \ T \ \mu^{8-2D} \
N \ S \ 2 (2-D) (b_1 - f_1) (4-D) f_2 \, \label{mia} \\
&& \nonumber \\
&=& {-e^5 T^4 N^{3/2} \over 64 \pi^3 \sqrt{3}}  + O(D-4) \, .
\label{ulala}
\end{eqnarray}
Pleasantly, while the individual expressions
$G_{15}$ and $G_{25}$
require the evaluation of some two-loop integrals, their sum
(\ref{mia}) depends  only on the trivial one-loop integrals
$b_n$ and $f_n$,
defined previously,  and  $S = \int {(dx) \over x^2 (x^2 +1)}$.
The final result as $D \rightarrow 4$ in (\ref{ulala})
is finite as
required since
the contribution of $e^5$ diagrams coming from
electron- wavefunction and vertex
renormalsations cancels as in the $e^4$ calculation
\cite{CP} because
of the Ward identity $Z_1 =Z_2$, and the restriction
to massless electrons.

The order $e^4 N^2$ diagram is shown in Fig.1. Inserting
self-energy
subdiagrams along the
photon lines produces the equivalent set of diagrams
shown in Fig. 2.
Power counting now indicates  that
only if the self-energies in Fig.2 are $\Pi_{00}(0,q)$ can one
obtain contributions of lower
order than $e^6$. These insertions sum to
\begin{eqnarray}
 - {T\over 2}\int (dq) \left[ \ln \left(1+{\Pi_{00}(0,q)
\over q^2}\right) - {\Pi_{00}(0,q)\over q^2}
+ {1 \over 2} \left({\Pi_{00}(0,q)\over q^2}\right)^2 \right].
\label{yuk}
\end{eqnarray}
However (\ref{yuk}) contains as a leading contribution
part of the
lower order  $e^3$ piece
due to $\Pi_{00}(00)=m^2$. Subtracting  from (\ref{yuk})
the same
expression but with $\Pi_{00}(0,q)$ replaced by $m^2$ gives,
upon using (\ref{burro}),
\begin{eqnarray}
 - {T\over 2}\int (dq) \left[ \ln \left(1+{e^2 q^2 h(q)
\over q^2 +
m^2}\right) -{e^2 h(q)}
+ {e^2 h(q)  \over 2 q^2} \left(2 m^2 + e^2 q^2 h(q)
\right) \right].
\label{soyuk}
\end{eqnarray}
This contains all terms of order $e^5 N^{5/2}$ but also many
 subleading terms.
Again the scaling $\vec{q}=m \vec{x} $ helps to identify
the exact $e^5$ pieces.
Dropping subleading terms one finds
\begin{eqnarray}
G_{35} &=&- {T e^2 m^{D-1} \over 2}\int (dx) \ x^2 \ h(0)
\left( {1 \over x^2 + 1} - {1 \over x^2}  + {1 \over x^4} \right)
\,\label{rose} \\
&& \nonumber \\
& \rightarrow & { -e^5 T^4 N^{5/2} \over  8 \pi \sqrt{27}} \
{\gamma-1 + \ln(4/\pi) \over 12  \pi^2}\ +
\ { e^5 T^4 N^{5/2} \over 8 \pi \sqrt{27}} \ {\ln(T/\mu) \over 6 \pi^2}
 \, .
 \label{chiquita}
\end{eqnarray}

The expression  (\ref{rose}) has a simple interpretation
and could have
been written down
by inspecting the diagrams in Fig. 2 :
It is the subleading
contribution from
the sum of infinite diagrams when, for each diagram,
the subleading $e^2q^2$ piece is
kept of exactly one self-energy while the
leading $e^2 T^2$ pieces are
taken from the remaining self-energies.
The $ e^5 N^{5/2} \ln (T/\mu)$ term in the
answer (\ref{chiquita}) is
a remnant
of wavefunction renormalisation. This
logarithm cancels
against a similar term that is generated
from the lower order $e^3 N^{3/2}$ plasmon term when the
pressure is
written in terms of the
temperature dependent coupling
\begin{eqnarray}
e^2(T)=e^2\left( 1+ {e^2 N \over 6 \pi^2}
\ln {T\over \mu}\right) + O (e^6) \ . \label{run}
\end{eqnarray}
Alternatively one may also
eliminate such logarithms by simply choosing $\mu =T$.

The fine-structure constant at temperature
$T$ is $\alpha(T)=e^2(T)/4\pi$.
Defining $g^2 = \alpha(T) N / \pi$,  the pressure of QED with
$N$ massless Dirac fermions at nonzero temperature, $T$,
then follows from
Refs. \cite{AP,CP} and eqns.(\ref{ulala}), (\ref{chiquita})
and  (\ref{run}),

\begin{eqnarray}
P\over T^4 &=& a_0 + g^2 a_2 + g^3 a_3 + g^4 (a_4 + b_4 / N) +
g^5 (a_5 + b_5 / N) + O(g^6) \, ,
\end{eqnarray}
with
\begin{eqnarray}
a_0 &=& {\pi^2\over 45} \ (1 +{7\over 4}N) \, , \\
a_2 &=& - {5\pi^2\over 72} \, ,\\
a_3 &=&  {2 \pi^2 \over 9 \sqrt{3} } \, , \\
a_4 &=& - 0.8216 \pm 10^{-4} \, , \\
b_4 &=&  1.2456 \pm 1.5 \times 10^{-4} \, , \\
a_5 &=&  { \pi^2 [1- \gamma - \ln(4 / \pi)] \over 9 \sqrt{3}} \
= \ 0.11473... \; , \\
b_5 &=&   {- \pi^2 \over 2 \sqrt{3} } \; .
\end{eqnarray}

Some observations are made on the
short perturbation series above.
For $N=1$ the fourth order ($g^4$) coefficient
$(a_4 +b_4)=0.4240 \pm 0.00025 $ is positive\footnote{This value
may be very closely approximated by
$3 \sqrt{2}/10 = 0.42426...$; also
note that $a_4 \approx - \pi^2/12 = - 0.8224..$.} while
the fifth order coefficient $(a_5 +b_5) = -2.734...$
is negative.
Thus there is a sign
alternation separately within  the even orders $(g^0, g^2, g^4)$
and the odd orders
$(g^3 , g^5)$.
On the other hand if one considers a double expansion in
$g$ and $1/N$
then the $(1/N)^0$
series \{${a_n}$\} shows sign alternation and so does
the $(1/N)$
series \{${b_n}$\}, at least for the
available terms. I do not know if this empirical regularity
is suggestive of a general result.

In $T=0$ QED the large order behaviour of perturbation theory
with respect to the expansion parameter $e^2$ was studied
many years ago
\cite{BIZP,GZ}. By contrast,
at $T \neq 0$, resumming the IR divergences
creates an expansion in
$\sqrt{e^2}$ and to my knowledge an asymptotic analysis
in this case is lacking.
Some information about the perturbation series can be obtained
in the
large $N$ expansion.
To leading order in $1/N$ (the \{$a_n$\} series),
the diagrams which
contribute are precisely those in Figs.1,2 (plus the order
$e^2$ diagram
not shown) with  {\it full} one-loop self-energy insertions.
Consider now only the odd terms $\{a_{2n+1} ,  n=1, 2,... \}$,
these are the plasmon
($e^{2n+1}$)
contributions in the large $N$ limit. From power counting
as before one
deduces that {\it all} the $a_{2n+1}$ terms are contained in the
expression
\begin{eqnarray}
 - {T\over 2}\int {d^3 q \over (2 \pi)^3 }
\left[ \ln \left(1+{\Pi_{00}(0,q) \over q^2}\right) -
{\Pi_{00}(0,q)\over q^2}  \right].
\label{hoolahoop}
\end{eqnarray}
Here $\Pi_{00}(0,q)$ is the finite one-loop self-energy
of eq.(\ref{burro})
at $D=4$ and renormalisation scale $\mu=T$.
A systematic expansion of (\ref{hoolahoop}) in $e$
determines the series
\{$a_{2n+1}$\} (parts of $a_{2n}$ are also contained
in the expansion) and it
might be possible to determine $a_{2n+1}$ in closed form
for any $n$.

The full $e^5$ calculation was simple because of the
factorisation of the
diagrams $G_1$ and $G_2$. In general, let the gauge-invariant
contribution to the pressure at order $e^{2n+2} \ (n \ge 1)$,
from diagrams with {\it one}-fermion loop, be $P^{1F}_{2n+2}$.
Then the order $P^{1F}_{2n+3}$
contribution is obtained by dressing and using
the factorisation property :

\begin{eqnarray}
P^{1F}_{2n+3} &=& \int {(dq) \ m^2 \over q^2 (q^2 + m ^2) } \
{ \delta P^{1F}_{2n+2} \over
\delta D_{00}(0,q \rightarrow 0)}\, V
\, , \\
& &\nonumber \\
&=& {eT N^{1/2} \over 4 \pi \sqrt{3}} \
{\delta P^{1F}_{2n+2} \over \delta D_{00}(0)} \ V \, , \\
&& \nonumber \\
&=& {e T^2  N^{1/2}\over 8 \pi \sqrt{3}} \
\hat{\Pi}_{00}^{1F; 2n+2} (0,0) \, ,
\label{zoo} \\
&& \nonumber \\
&=& {e^3 T^2 N^{1/2} \over 8 \pi \sqrt{3}} \
{\partial^{2} P_{2n}^{1F} \over \partial
\mu^2} |_{\mu=0} \; \; \; \, , n \ge 1 \, .
\label{spot}
\end{eqnarray}
Equation (\ref{zoo}) follows from the
earlier equation upon using the
relation \cite{book} $ \left({\delta P \over \delta D_{\mu \nu}}
\right)_{IPI} = { T \ \hat{\Pi}^{\mu \nu} / 2 V }$ ,
where $V=$volume, $IPI$ refers to one-particle-irreducible
and $\hat{\Pi}$ is the full self-energy. The final equation
(\ref{spot})
then follows from the identity \cite{book}
$\hat{\Pi}_{00}(00) = e^2 { \partial P^2 / \partial \mu^2}$.
The result
(\ref{spot}) is that mentioned in the abstract.
It is valid\footnote{An alternative derivation using the
ring-summation formula (\cite{book}),
and the extension to massive electrons at nonzero chemical potential
will be presented elsewhere.} for the case of
$N$ massless electrons at zero chemical potential $\mu$.
For the case $n=1$
one may verify that (\ref{spot}) is satisfied by computing the
right-hand-side from known two-loop results \cite{AP,book} and
comparing it with the left-hand-side given by (\ref{ulala}).

Eq.(\ref{spot}) suggests that the  $e^7$ contribution
should be calculable
because knowledge of three-loop
diagrams is at hand \cite{FST,CP} (diagrams with more
than one fermion loop
are not covered by (\ref{spot}) but they are easier
and are handled
as for the case Fig.1 discussed above). If extra
cancellations occur,
as in the $e^5$ computation,
 then the $e^7$ calculation might reduce to a two-loop
calculation.
Note that the factorisation is successful because fermions
in imaginary time have IR safe propagators. If one attempted
to calculate
the $\lambda^5$ contribution to
the pressure of $\lambda^2 \phi^4$ theory, the diagram
in Fig. 4 is involved.
Now the exact factorisation fails
because the two-loop subdiagram has a logarithmic singularity
coming from the zero modes
of the bosonic propagators. Nevertheless the calculation
in this case is
still feasible  because
only the leading $ T^2 ( \ln k +$ constant) piece
of the two-loop
self-energy subdiagram (with inflowing
momentum $(0,k)$) is needed to
determine the order $T^4 \lambda^5 (\ln \lambda +$ constant)
contribution .

Calculations similar to those discussed in \cite{CP} and
here have been performed to obtain directly
the  screening masses in hot QED to high-order \cite{BIP},
and future
applications to quantum chromodynamics \cite{KT} are envisaged.
One speculative application for  QED will be mentioned :
 there has been much discussion in the literature
(see, e.g., \cite{AZ} and references therein)
concerning a possible strong-coupling phase (at $T=0$)
of QED with an ultraviolet-stable fixed point. Usual
perturbative arguments
indicate that at exponentially
high temperatures the QED coupling will be strong. If
a temperature driven transition to a stable nonperturbative
phase
is possible, then a resummed high-order perturbation
series \cite{GZ}
for hot QED might be useful for some studies.

In conclusion,
some of the questions raised in the last three paragraphs
are left for future
investigations \cite{CP2}.\\

\smallskip
\smallskip
Acknowledgements : I thank C.Corian\`{o}, J.P. Blaizot,
R.Pisarski,
T. Hudson and S. Mennecier for helpful discussions.

\pagebreak
\noindent{\underline{\Large Figure Captions}}
\smallskip

\underline{Fig.1}:\\ The order $e^4 N^2$ diagram.
It has an infrared
singularity which contributes to the lower ($e^3$) order.
In Figs.1-3 the wavy
line is the photon.

\underline{Fig.2}: \\ Diagrams generated from Fig.1 by
photon-polarisation
insertions.

\underline{Fig.3}:\\  The order $e^4 N$ diagrams
which produce $e^5 N^{3/2}$
terms when the photons are dressed.

\underline{Fig.4}:\\
The nontrivial three-loop diagram of $\lambda^2 \phi^4$ theory
studied in \cite{FST}. It will contribute a
$\lambda^5 \ln \lambda$ piece
to the free energy when dressed.

\pagebreak


\newpage
\begin{thebibliography}{99}
\bibitem{CP} C. Corian\`{o} and R.R. Parwani, preprint
ANL-HEP-PR-94-02 (Argonne), SPhT/94-054 (Saclay),
hep-ph/ 9405343.
\bibitem{AP} I.A. Akhiezer and S.V. Peletminskii,
Sov. Phys. JETP 11 (1960) 1316.
\bibitem{FMB} B.A. Freedman and L.D. McLerran,
Phys. Rev. D16 (1977) 1130, 1147, 1169; \\
 V. Baluni, Phys. Rev. D17 (1978) 2092.
\bibitem{FST} J. Frenkel, A.V. Saa and J.C. Taylor, Phys.
Rev. D46 (1992) 3670.
\bibitem{GB} M. Gell-Mann and K.A. Brueckner,
Phys. Rev. 106 (1957) 364.
\bibitem{BIZP} R. Balian, C. Itzykson, J.B. Zuber and G. Parisi,
Phys. Rev. D17, (1978) 1041.
\bibitem{GZ} ``Large Order Behaviour of Perturbation Theory'', \\
J.C. Le Guillou and J. Zinn-Justin eds. (North Holland, 1990).
\bibitem{book} J. Kapusta, ``Finite Temperature Field Theory''
 (Cambridge University Press, 1989).
\bibitem{BIP} J.P. Blaizot, E. Iancu and R.R. Parwani,
manuscript in preparation.
\bibitem{KT} J.I. Kapusta, Nucl. Phys. B148 (1979) 461; \\
T. Toimela, Phys. Lett. 124B, (1983) 407.
\bibitem{AZ} M. Awada and D. Zoller, Phys. Lett. B299,
(1993) 151; \\
K.-I. Kondo, T.Iizuka, E.Tanaka and T.Ebihara,
Phys. Lett. B325 (1994) 423;\\
 A.S. Goldhaber, H.N. Li and R.R. Parwani, preprint SPhT/93-047,
hep-th/9305007, submitted to Phys. Rev. D.
\bibitem{CP2} C. Corian\`{o} and R.R. Parwani, in progress.
\end{thebibliography}
\end {document}